\RequirePackage{fix-cm}
\documentclass[twocolumn]{svjour3}          % twocolumn
\usepackage{graphicx}
\usepackage{amssymb}
\usepackage{amsfonts}
\usepackage{amsmath}
\usepackage{latexsym}
 \journalname{Foundations of Physics}
\begin{document}

\title{Testing a Quantum Inequality \\with a Meta-analysis of Data for Squeezed Light%\thanks{Grants or other notes
}

%\titlerunning{Testing a Quantum Inequality%Short form of title}        % if too long for running head

\author{G. Jordan Maclay \and
        Eric W. Davis %etc.
}


\institute{G. Jordan Maclay \at
          Quantum Fields LLC, St. Charles, IL 60174 \\
             \email{jordanmaclay@quantumfields.com}           %  \\
           \and
           Eric W. Davis \at
             Institute for Advanced Studies at Austin, 11855 Research Blvd., Austin, TX
78759; Early Universe, Cosmology and Strings Group, Center for Astrophysics, Space Physics and Engineering Research, Baylor University, Waco, TX 76798\\
\email{EWDavis@earthtech.org}
}

\date{Received: date / Accepted: date}

\maketitle

\begin{abstract}
In quantum field theory, coherent states can be created that have negative
energy density, meaning it is below that of empty space, the free quantum
vacuum. If no restrictions existed regarding the concentration and permanence of negative energy regions, it might, for example, be possible to produce exotic phenomena such as Lorentzian traversable wormholes, warp drives, time machines, violations of the second law of thermodynamics, and naked singularities.  \ Quantum Inequalities (QIs) have been proposed that restrict the size
and duration of the regions of negative quantum vacuum energy that can be
accessed by observers. \ However, QIs generally are derived for situations in cosmology and are very difficult to test. \ Direct measurement of vacuum energy is difficult and to date no QI has been tested experimentally. \ We
test a proposed QI for squeezed light by a meta-analysis of published data obtained
from experiments with optical parametric amplifiers (OPA) and balanced homodyne
detection. Over the last three decades, researchers in quantum optics have been trying to maximize the squeezing of the quantum vacuum and have succeeded in reducing the variance in the quantum vacuum fluctuations to -15 dB. To apply the QI, a time sampling function is required.  In our meta-analysis different time sampling functions for the QI were examined, but in
all physically reasonable cases the QI is violated by much or all of the
measured data. This brings into question the basis for QI. \ Possible explanations are given for this surprising result.
%Insert your abstract here. Include keywords, PACS and mathematical
%subject classification numbers as needed.

%\pacs{42.50Lc - Vacuum fluctuations}
%\pacs{PACS. 42.50Dv - Nonclassical field states; squeezed, antibunched, and
%sub-Poissonian states; operational definitions of the %phase of the field;
%phase measurements.}
%\pacs{3.65Wj, 11.10.Ef, 42.50Lc, 42.50Dv }

\end{abstract}

\section{Introduction}
\label{intro}
In quantum field theory, the vacuum expectation value of the normally ordered or
renormalized energy density $\langle T_{oo}\rangle$ need not be positive.
\ For example, a superposition of a vacuum state (n=0) and a two photon state
(n=2), can have negative renormalized energy density with the proper choice of
coefficients. \ Squeezed light can have a negative energy density. From theory
and experiment, we know that static negative energy densities associated with
vacuum states are concentrated in narrow spatial regions, e. g., inside a
parallel plate Casimir cavity with small plate separation or in the region
near the Schwarzschild radius in the Boulware vacuum where the energy density
is everywhere negative as seen by static observers. There is no known way to directly measure vacuum energy density.  \ On the other hand, the
total energy of a system is believed to always be positive or zero. \ For
example, the sum of the mass energy of the plates plus the negative vacuum
energy inside the cavity is positive \cite{beckenstein}\cite{Visser}. \ The
classical energy conditions imply that an inertial observer who initially
encounters some negative energy density must encounter compensating positive
energy density at some arbitrary time in the future. \ Quantum Inequalities
(QIs) have been derived for the free vacuum quantum electromgnetic field, with no sources or
boundaries, which constrain the magnitude and duration of negative energy
densities relative to the energy density of an underlying reference vacuum
state. \ The QI places bounds on quantum violations of the classical energy
conditions \cite{ford}\cite{davies}. \ The QI is formulated as a mathematical
bound on the average of the quantum expectation value of a free field's
energy-momentum tensor in the vacuum state, where the average is taken along
an observer's timelike or null worldline using time sampling functions.
\ Contrary to the classical energy conditions, the QI dictates that the more
negative the energy density is in some time interval T, the shorter the
duration T of the interval, so that an inertial observer cannot encounter
arbitrarily large negative energy densities that last for arbitrarily long
time intervals. \ An inertial observer must encounter compensating positive
energy density no later than after a time T, which is inversely proportional
to the magnitude of the initial negative energy density.

In QI, restrictions are placed on the integral of the vacuum expectation value of the renormalized
energy density $<T_{oo}>$ multiplied by a sampling function. \ For the electromagnetic
field in flat space-time, with a normalized time sampling function of
$\ f(t)=(t_{o}/\pi)(1/(t^{2}+t_{o}^{2})),$ Ford has shown
\cite{Ford2}
\begin{equation}
\hat{\rho}\equiv\frac{t_{o}}{\pi}\int_{-\infty}^{+\infty}\frac{\langle
T_{oo}\rangle}{t^{2}+t_{o}^{2}}dt \geqslant -\frac{3}{16\pi^{2}}\frac{\hbar
c}{(ct_{o})^{4}}%
\end{equation}

To give a frame of reference, this can be compared to the vacuum energy density within
an ideal parallel plate Casimir cavity of separation $a$:%
\begin{equation}
\langle T_{oo}\rangle_{Cas}=-\frac{\pi^{2}}{720}\frac{\hbar c}{a^{4}}%
\end{equation}
The ratio of the numerical factors for the free field to the Casimir cavity is
1.4, so a negative energy density $\hat{\rho}$\ equal to that in a perfectly
conducting parallel plate cavity of spacing $a$ can exist no longer than for a
time $t_{o}\sim a/c$, about $3\times10^{-16}$ seconds for a typical experiment. 
\ As the sampling time $t_{o}$ increases, $\hat{\rho}$
rapidly goes to zero. \ (Note however, that as derived, the QIs do not apply
directly to the Casimir cavity since it has boundaries. Also, to test Eq. 1 experimentally, one must make an absolute measurement of the vacuum energy density,  
an experimental challenge for which no solution has yet been found. Some progress is due to Riek et al. who were able to directly probe the spectrum of squeezed vacuum fluctuations of the electric field in the multi-THz range using femtosecond laser pulses \cite{reik}.) 

If the laws of quantum field theory placed no restrictions on negative energy,
then it might be possible to produce surprising macroscopic effects such as
violations of the second law of thermodynamics, traversable wormholes, warp
drives, and possibly time machines \cite{Ford2}. \ QI appear to restrict these
violations of the second law \cite{ford3}\cite{ford5}. \ 

A quantum inequality has been derived for squeezed light by
Marecki  \cite{marecki1}. \ Squeezed light has a nonclassical distribution of
the quadrature components (typically phase and amplitude), which may be considered as the canonical momentum
and position components of an equivalent harmonic oscillator corresponding to
the frequency of the electromagnetic radiation being considered. \ Squeezed
states are routinely made in quantum optics experiments in the process of
parametric down conversion, in which an incident photon is converted in a
non-linear crystal to two entangled photons of the same frequency, which is
one half of that of the incident photon. \ The fluctuations of the electric
field in the squeezed light are locally lower than the vacuum fluctuations,
the so-called shot-noise level. \ There appears to be a limit to the amount of
squeezing relative to the free vacuum, which has been measured to be from -0.5
dB to the most recent value of -15 dB \cite{vah}. Detection of the squeezing relative to the free vacuum field is done using balanced homodyne detection (BHD). \ Marecki's QI predicts the maximum degree of squeezing in dB that is possible in terms of the fraction
of the cycle during which the variance in the electric field is less than that
of the free vacuum limit. Marecki developed the theoretical framework demonstrating the ability of BHD to quantify the vacuum fluctuations of the electric field in terms of vacuum expectation values of products of the electric field operators, the one- and two- point functions of arbitrary states of the electric field\cite{marecki3}\cite{marecki2}. He applied this theory to the measurement of negative Casimir energy densities using BHD. The corresponding experiments require the placement of photodiodes within Casimir cavities and have yet to be performed.

To date no QI has been tested experimentally. \ One of the reasons for this was noted by Marecki \cite{marecki1}: "As far as we know quantum field theoreticians do not
know that their inequalities may influence real experiments
nor are quantum opticians aware of the existence of such
inequalities."  Most of the quantum inequalities have been developed by quantum cosmologists or quantum field theorists, who are unaware of measurements of vacuum energy done by experts in quantum optics. This paper is the first attempt to bridge this gap and test a quantum inequality with published experimental data. It appears easiest to test the
QI for squeezed light because with balanced homodyne detection one measures the squeezing
relative to the free vacuum, which corresponds to the theoretical quantity described in the corresponding quantum inequality.
\ However, there may be some subtleties in the comparison because of
differences in the measurement protocols. Indeed, we find that the QI as given
is violated by\ most of the experimental data, yet all experimental data are
consistent with a theoretical model of the optical parametric amplifier (OPA) used to generate squeezed light. 

%Your text comes here. Separate text sections with
%\section{Section title}
%\label{sec:1}
\section{Quantum Inequality for Squeezed Light}
\label{sec:1}
The QI for squeezed light gives a minimum value for the time sampled magnitude of the variance $<\Delta>_{A}$ of the quantized electromagnetic field for a state A where

\begin{multline}
<\Delta>_A
\equiv \\ \int_{-\infty}^{+\infty}f(t)dt(<E^{2}(x,t)>_A-<E^{2}(x,t)>_{vac})%
\end{multline}

A simplified version of Marecki's derivation is given in Appendix A.  His key result is \cite{marecki1}
\begin{multline}
<\Delta>_{A}\quad   \geq \\ \frac{-2}{(2\pi)^{2}}\int_{0}^{\infty}d\omega\int
d^{3}p\mu_{p}^{2}\omega_{p}|(f^{1/2})_{FT}(\omega+\omega_{p})|^{2} \label{res}%
\end{multline}
where $\omega^{2}_{p}=p_{1}^2+p_{2}^2+p_{3}^2$.

 \ The minimum value
of the variance $<\Delta>_{A}$ for a state A is determined by the time window function $f(t)$, specifically by $|(f^{1/2})_{FT}|^{2}$
the magnitude squared of the Fourier transform of the square root of the window function $f(t)$. In order to insure convergence of the integral, a spectral function $\mu_{p}=$ $\mu(\omega_{p}-\omega_{0})$, a function of $\omega_{p}$ that is strongly peaked at $\omega_{p}=\omega_{0}$, must be included.  This term reflects the frequency response of
the apparatus measuring the variance. The
result Eq. 4 is similar in spirit to that of other researchers in that it involves
the Fourier transform of the time window \cite{pfen}.\ There is no proof
that Eq. \ref{res} represents the greatest lower bound for $<\Delta>_{A}.$

In formulations of other Quantum Inequalities, other features of the time
window, such as the second derivative, determine the minimum average energy
over the time sampling \cite{Ford2}. \ In all formulations of Quantum
Inequalities to date, the window function determines the minimum energy
values. This may seem counter intuitive. \ In all cases, these formulations assume a free plane-wave electromagnetic field
without sources or \\
boundaries. We have found,
like others, that the specific properties of the window function are very
important \cite{davies}. \ Only in Marecki's calculations of the\ QI does a
spectral function $\mu_{p}$ appear. \ This\ may be a problem because the
Fourier transform of the time window $f(t)$ implies a certain frequency
response of the apparatus, and this may conflict with the independent
requirements for the function $\mu_{p}.$
The quantity that is generally measured in experiments is the $Log_{10}$ of the
variance for some state A relative to the variance of the free vacuum:  
\begin{equation}
R_{expt}=10Log_{10}\left(  \frac{<\Delta>_{A}+<E^{2}>_{vac}}{<E^{2}>_{vac}}\right)
\end{equation}%
According to Marecki, the measured squeezing in dB
must exceed in numeric value $R$, where%
\begin{multline}
R=10Log_{10}\Bigg[\\ \frac{\frac{1}{(2\pi)^{3}}\int d^{3}p\mu_{p}^{2}\omega
_{p}(1-\int_{0}^{\infty}d\omega(4\pi|(f^{1/2})_{FT}(\omega+\omega_{p})|^{2}%
)}{\frac{1}{(2\pi)^{3}}\int(\mu_{p}^{2}\omega_{p})d^{3}p}\Bigg]
\end{multline}

%\subsection{Subsection title}
%\label{sec:2}
\subsection{Evaluation of R for Specific Time Sampling Functions}
\label{sec:2}
For a Gaussian
\begin{equation}
f(t)=\frac{1}{t_{0}\sqrt{2\pi}}e^{-t^{2}/2t_{0}^{2}}%
\end{equation}
we obtain%
\begin{equation}
R=10Log_{10}\left[  -\frac{\int d^{3}p\mu_{p}^{2}\omega_{p}Erf[\sqrt{2}%
t_{o}\omega_{p}]}{\int d^{3}p\mu_{p}^{2}\omega_{p}}\right]
\end{equation}
If $\mu_{p}$ is a function of $\omega_{p}$ sharply peaked at $\omega_{0}$, with width
$\delta\omega<<\omega_{0}$, then to a good approximation%
\begin{equation}
R(\omega_{o}t_{o})=10Log_{10}[Erf(\sqrt{2}\omega_{o}t_{o})]
\end{equation}
(Marecki has an additional factor of 2 in the Erf function, which we do not
get). \ As a check on the role of the frequency windows $\mu_{p}$, we can do
all the integrations in $R$ for a Gaussian frequency function, and we get an
additional factor of $\omega_{0}^{3}\delta\omega$ for the $\omega_{p}$
integration in the numerator and in the denominator. \ These factors cancel,
giving to lowest order in $\delta\omega$, the result quoted above. \ On the
other hand, if we do not introduce a frequency function, we find that
$R=10Log_{10}[1]$ $=0,$ indicating that no squeezing is possible. \ In other
words, a frequency function is required to get reasonable results; however, as
we have noted the frequency function may not be consistent with the time
window. \

We can also compute $R$ for a squared Lorentzian time sampling function (an
ordinary Lorentzian does not give well-behaved integrals):%
\begin{equation}
f(t)=\frac{2}{\pi}\frac{t_{0}^{3}}{(t^{2}+t_{o}^{2})^{2}}%
\end{equation}
We find
\begin{equation}
R(\omega_{o}t_{o})=10Log_{10}(1-e^{-2\omega_{o}t_{o}})
\end{equation}
assuming \ $\mu_{p}$ is strongly peaked at $\omega_{o}.$ \ The equation
behaves similarly to the one for the Gaussian time function.

We can compute $R$ for a square window function $f(t)$ of width $\Delta T$ 
with perfectly sharp corners and using a frequency function $\mu_p$. We find that
we always get perfect squeezing, $R=10Log_{10}0$, with no dependence on $\Delta T$.
Although a perfectly sharp window is not physically possible, and is
mathematically unstable, one still wonders about the meaning of this result.
A sharp window allows one to do a perfect measurement (at least in
principle) in which only regions of perfect squeezing are measured, and one
can avoid the regions with partial or antisqueezing.

We have also evaluated the variance for a symmetric trapezoidal window with a
center region $T_{S}$ long and sloping sides that are each $nT_{s}$ long,
normalized to 1.

\section{Production of Squeezed Light Using Optical Parametric Amplification}

\begin{figure}[ptb]%
\centering
\includegraphics[height=2.0911in, width=3.3035in]{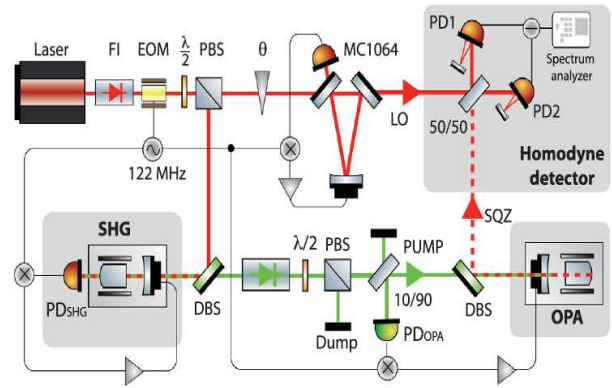}
\caption{Schematic of experimental setup. Squeezed vacuum states of light SQZ
at a wavelength of 1064 nm were generated in a double resonant, type I optical
parametric amplifier (OPA) operated below threshold. \ SHG: second-harmonic
generator; PBS: polarizing beam splitter; DBS: dichoric beam splitter;
LO: local oscillator; PD: photodiode; MC1064: three mirror ring cavity for
spatiotemporal mode cleaning; EOM: electro-optical modulator; FI: Faraday
isolator. The phase shifter for the relative phase $\theta$ between SQZ and LO
was a piezoelectric actuated mirror\cite{vah}. }%
\label{1}%
\end{figure}

A model for an optical parametric amplifier (OPA) with balanced homodyne detection (BHD) predicts the relative variance S
in the quadrature components of the vacuum electromagnetic field for a state
A:%
\begin{equation}
S=\frac{\langle E^{2}\rangle_{A}}{\langle
E^{2}\rangle_{vac}}%
\end{equation}
\ The model \cite{gar,collandwalls,polzit} predicts that%
\begin{multline}
S(\theta,x,\omega)=1+4\beta x\bigg[  \frac{\operatorname{cos^{2}\theta}}{(1-x)^{2}%
+(\omega/\gamma)^{2}}  -\\ \frac{\operatorname{sin^{2}\theta}}{(1+x)^{2}%
+(\omega/\gamma)^{2}} \bigg]
\end{multline}
where $x =P/P_{th}$ is the ratio of the laser power to the power at threshold $(0<x<1)$, $\beta$ is the optical efficiency, $\theta$ is the phase difference
between the local oscillator field (LO) and the vacuum field, $\omega$ is the
sideband angular frequency of measurement by a spectrum analyzer, $\gamma$ is
the halfwidth or cavity decay rate [$\gamma=c(T+L)/l$ where $c=$ speed of
light, $T=$transmissivity of coupling mirror, $L=$round trip loss, $l=$round trip length]. The model has been parameterized so the squeezing is a maximum
at $\omega=0.$ \ Generally, the squeezing is given in terms of dB:%
\begin{equation}
R=10Log_{10}S(\theta,x,\omega)
\end{equation}
To clarify the physical basis of the model and derive equations relating it to
the QI, we briefly review the OPA model and experimental results. \ In recent
experiments, values for the full width $2\gamma/2\pi$ ranges from $9$
MHz to $84$ MHz. \ Measurement frequencies $\omega$ are typically about $1$
MHz to, at most, $8$ MHz, $0.9<\beta<0.99,$ and laser wavelengths vary
from about $795$ nm to $1064$ nm. \ In the measurement range, $(\omega
/\gamma)$ varies from about 0 to, at most, 1. Figure \ref{1} shows a
recent experimental arrangement \cite{vah}.  A CW laser with frequency $\omega_{LO}$ encounters a polarizing beam splitter PBS, one beam going to a second harmonic generator SHG, the other beam which serves as the local oscillator LO goes to a 50-50 beam splitter in the balanced homodyne detector BHD. The beam leaving the SHG, with frequency $2\omega_{LO}$, goes to the optical parametric amplifier OPA.  The OPA is operated below
threshold and is composed of a cavity with a nonlinear crystal that is fully
reflective at one end and a partially reflective mirror at the other end.
The OPA non-linear crystal is driven by the output of a frequency-doubled
laser SHG. The crystal has a small probability of producing two photons of
the same frequency $\omega_{LO}$ (half the driving frequency) by degenerate parametric
down conversion. \ Detection is by balanced homodyne detection in which the
difference in photodetector current PD1-PD2 is measured for components of the
squeezed vacuum SQZ and the laser LO that have interfered
at a 50-50 beam splitter. The difference current is analyzed by a spectrum
analyzer, typically with a measurement bandwidth of about 100 kHz to 500 kHz. %
%TCIMACRO{\FRAME{ftbpFU}{3.4835in}{2.0911in}{0pt}{\Qcb{Schematic of
%experimental setup. Squeezed vacuum states of light SQZ at a wavelength of
%1064 nm were generated in a double resonant, type I optical parametric
%amplifier (OPA) operated below threshold. \ SHG: second-harmonic generator;
%PBS: polarizing beam splitter; DBS: dichoric beam splitter; LO:local
%oscillator; PD: photodiode; MC1064: three mirror ring cavity for
%spatiotemporal mode cleaning; EOM: electro-optical modulator; FI: Faraday
%isolator. The phase shifter for the relative phase $\theta$ between SQZ and LO
%was a piezoelectric actuated mirror\cite{vah}. }}{\Qlb{1}}{Figure
%1}{\special{ language "Scientific Word";  type "GRAPHIC";  display "USEDEF";
%valid_file "T";  width 3.4835in;  height 2.0911in;  depth 0pt;
%original-width 8.6481in;  original-height 4.5498in;  cropleft "0";
%tempfilename 'P1ADMA0D.bmp';tempfile-properties "XPR";}}}%
%BeginExpansion
%EndExpansion

\begin{figure}[ptb]%
\centering
\includegraphics[
trim=0.000000in 0.000000in 0.000000in -0.318599in,
height=2.4379in,
width=3.3126in
]%
{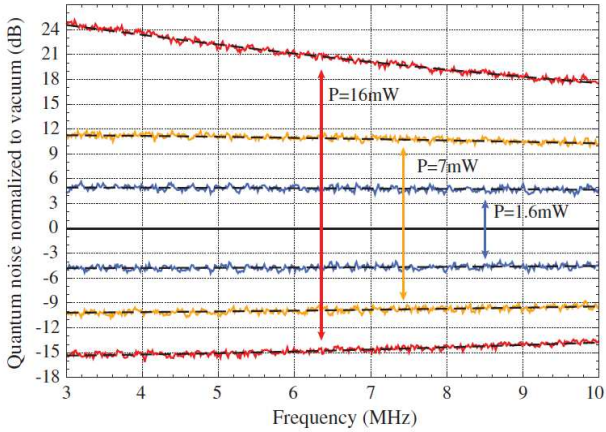}%
\caption{Squeezing in $dB=10Log_{10}R$ as a function of $Power=P_{th} x$, $P_{th}= 16.2$ $ 
mW$, and measurement frequency $\omega. $ Theoretical curves are shown as the narrow dashed
lines with 2$\gamma/2\pi$ =84 MHz and $\beta=0.975.$ The decrease in noise
with increase in frequency is due to the term ($\omega/\gamma)^{2}$ in Eq.
13 \cite{vah}.}%
\label{sqx}%
\end{figure}

Data on a squeezed vacuum taken from the apparatus illustrated are shown in
Figure \ref{sqx} \cite{vah}. Fits based on the maximum and minimum values of $S$ from Eq. 13 are shown in dashed
lines for three power settings.%
%TCIMACRO{\FRAME{ftbpFU}{3.4126in}{2.4379in}{0pt}{\Qcb{Squeezing in
%dB=10LogR$,$ as a function of Power=Pth $x,$ Pth= 16.2 mW, and frequency
%$\omega. $ Theoretical curves are shown as the narrow dashed lines, with
%2$\gamma/2\pi$ =84 MHz, and $\beta=0.975.$ The decrease in noise with increase
%in frequency is due to the term ($\omega/\gamma)^{2}$ in Eq \ref{2xx}%
%.\cite{vah}}}{\Qlb{sqx}}{Figure 2a}{\special{ language "Scientific Word";
%type "GRAPHIC";  display "USEDEF";  valid_file "T";  width 3.4126in;
%height 2.4379in;  depth 0pt;  original-width 8.6931in;
%original-height 6.0113in;  cropleft "0";  croptop "1.0530";  cropright "1";
%tempfilename 'jordan/P1ADMA0E.bmp';tempfile-properties "XPR";}}}%
%BeginExpansion
%EndExpansion%
%TCIMACRO{\FRAME{ftbpFU}{3.3883in}{2.4474in}{0pt}{\Qcb{Squeezing dB as a
%function of phase difference $\theta,$ measured here by the time to move a
%mirror. \ Curve a: noise level with all inputs blanked; Curve b: phase is
%locked to the squeezed quadrature ($\theta=\pi/2)$; Curve c: phase is locked
%to the antisqueezed quadrature ($\theta=0)$; Curve d: the phase is scanned.
%\ The fraction of the period that the squeezing is below zero equals
%FT.\ \cite{vah}}}{\Qlb{3}}{Figure}{\special{ language "Scientific Word";
%type "GRAPHIC";  maintain-aspect-ratio TRUE;  display "USEDEF";
%valid_file "T";  width 3.3883in;  height 2.4474in;  depth 0pt;
%original-width 8.7908in;  original-height 6.3399in;  cropleft "0";
%tempfilename 'P1ADMA0F.bmp';tempfile-properties "XPR";}}}%
%BeginExpansion
\begin{figure}[ptb]%
\centering
\includegraphics[
height=2.4474in,
width=3.2883in
]%
{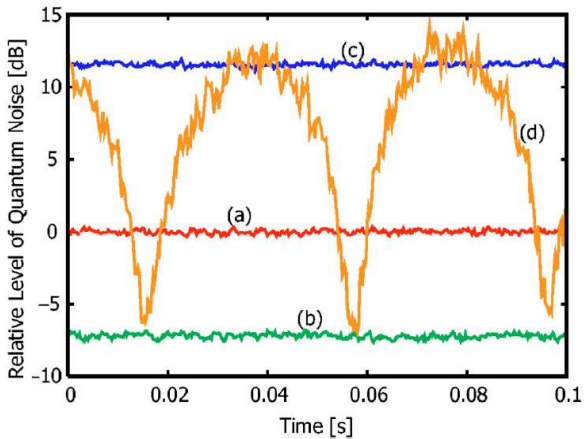}%
\caption{Squeezing dB as a function of phase difference $\theta$ measured
here by the time to move a mirror. \ Curve a: noise level with all inputs
blanked; Curve b: phase is locked to the squeezed quadrature ($\theta=\pi/2)$;
Curve c: phase is locked to the antisqueezed quadrature ($\theta=0)$; Curve d:
the phase is scanned. \ The fraction of the period that the squeezing is below
zero equals $F_{T}$ \cite{vah}.}%
\label{3}%
\end{figure}
In Figure 3, the vacuum squeezing is presented as a function of the phase
difference between the LO and the squeezed vacuum SQZ \cite{vah}. \ In this
particular experiment the mirror was vibrated periodically and the abscissa
given in time rather than angle, but these methods are equivalent. \ The first
minimum would correspond to the antisqueezed quadrature $\theta=\pi/2$
radians, the second to $\pi/2$ +$\pi$ radians \cite{suzuki}. The fraction of
the period for which the squeezing is negative equals $F_{T},$ which is an
indicator of the squeezing. From measuring the graph (using curves d and a),
one finds that $F_{T}$ is about 0.14.

If $S(\theta,x,\omega)<1$, then the variance or noise of this
quadrature component is less than that of the free vacuum and is squeezed.
This implies that the other quadrature component $R(\theta+\pi/2,x,\omega)>1$
is antisqueezed. The minimum value of S for squeezing occurs for
$\theta=\pi/2$ and equals
\begin{equation}
S_{-}(x,\omega)=1-\frac{4\beta x}{(1+x)^{2}+(\omega/\gamma)^{2}}%
\end{equation}
and the maximum antisqueezing occurs for $\theta=0$ or $\pi$ and equals 
\begin{equation}
S_{+}(x,\omega)=1+\frac{4\beta x}{(1-x)^{2}+(\omega/\gamma)^{2}}%
\end{equation}

\begin{figure}[ptb]%
\centering
\includegraphics[
height=2.034in,
width=3.2534in
]%
{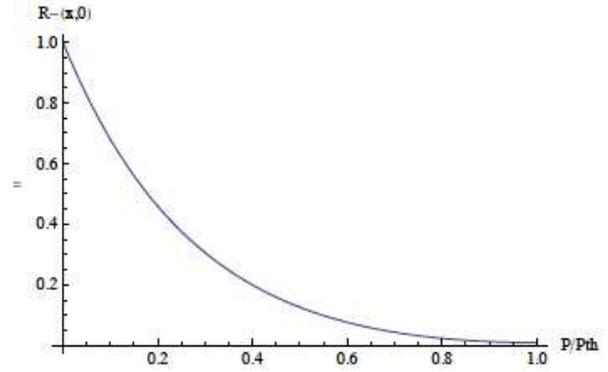}%
\caption{The maximum squeezing R$_{-}=$ S$_{-}(x,\omega)$ as a function of x.
\ When S$_{-}(x,\omega)<1$ there is squeezing below the normal
quantum limit. We assume $\beta=1$ and $\omega/\gamma$ is negligible. }%
\label{4}%
\end{figure}

The maximum possible squeezing $S_{-}(x,0)$ is shown as a function of $x$ in
Figure \ref{4}. %
%TCIMACRO{\FRAME{ftbpFU}{3.2534in}{2.034in}{0pt}{\Qcb{The maximum squeezing
%R$_{-}=$ S$_{-}(x,\omega)$ as a function of x. \ When S$_{-}(x,\omega)<1$ there is squeezing below the normal quantum limit. We assume
%$\beta=1$ and $\omega/\gamma$ is negligible. }}{\Qlb{4}}{Figure}%
%{\special{ language "Scientific Word";  type "GRAPHIC";  display "USEDEF";
%valid_file "T";  width 3.2534in;  height 2.034in;  depth 0pt;
%original-width 4.1105in;  original-height 2.4621in;  cropleft "0";
%tempfilename 'P1ADMA0G.bmp';tempfile-properties "XPR";}}}%
%BeginExpansion
%EndExpansion
The frequency spectrum of squeezing $1-S_{-}(x,f=\omega/\gamma)$ \ is
Lorentzian with a halfwidth of $\gamma.$ \ The product of the maximum and
minimum variances is
\begin{multline}
S_{-}(x,\omega)\ast S_{+}(x,\omega)=1-\\ \frac{16\beta(1-\beta)x^{2}}{\left[
(1+x)^{2}+(\omega/\gamma)^{2}\right]  \left[  (1-x)^{2}+(\omega/\gamma
)^{2}\right]  }%
\end{multline}

For an ideal optical system with no losses $\beta=1$ and the product is 1, as
it must be according to the Heisenberg Uncertainty Principle.

\begin{figure}[ptb]%
\centering
\includegraphics[
height=1.8248in,
width=3.0104in
]%
{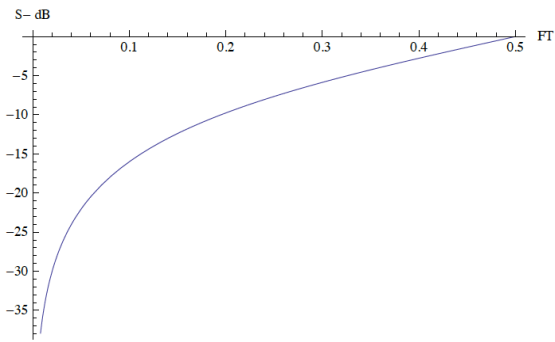}%
\caption{$10Log_{10}S_{-}(x,\omega)$ vs. $F_{T}(x,\omega)$ for an ideal OPA.}%
\label{db}%
\end{figure}

For comparison to the quantum inequality, we need to know the angular interval
$\Delta\theta$\ over which the light is squeezed. The light is squeezed if the
term in brackets in Eq. 13 is negative, which implies
\begin{equation}
\frac{S_{+}(x,\omega)-1}{1-S_{-}(x,\omega)}<\tan^{2}\theta
\end{equation}
It follows that $F_{T}=\Delta\theta/\pi$, which is the fraction of the period
during which the light is squeezed, is given by%
\begin{equation}
F_{T}(x,\omega)=1-\frac{2}{\pi}\tan^{-1}\sqrt{\frac{S_{+}(x,\omega)-1}%
{1-S_{-}(x,\omega)}} \label{ftgen}%
\end{equation}
For the special case of an ideal OPA, we can substitute $S_{+}=1/S_{-}$ to get%
\begin{equation}
F_{T}(x,\omega)=1-\frac{2}{\pi}\tan^{-1}\sqrt{\frac{1}{S_{-}(x,\omega)}}
\label{ftideal}%
\end{equation}
which can be solved for $S_{-}$ to obtain%
\begin{equation}
S_{-}(x,\omega)=\tan^{2}[F_{T}(x,\omega)\frac{\pi}{2}]
\end{equation}
which is valid for $0<F_{T}<0.5$. A plot of $R =\\ 10Log_{10}S_{-}(x,\omega)$ as
a function of $F_{T}(x,\omega)$\ for an ideal OPA is shown in Figure \ref{db}.
We would not expect experimental points to display squeezing greater than the amount allowed for the ideal OPA. Since most OPA measurements were not ideal, we used the general formula Eq.
\ref{ftgen} for $F_{T}(x,\omega)$ to reduce data. \

The maximum fraction of time in a period during which squeezing can occur is
$F_{T}=1/2$, and this only occurs when $x$ approaches zero, so the amount
of squeezing is slight.
%TCIMACRO{\FRAME{ftbpFU}{3.0104in}{1.8248in}{0pt}{\Qcb{$10LogS_{-}(x,\omega)$
%vs $F_{T}(x,\omega)$ for an ideal OPA.}}{\Qlb{db}}{Figure}%
%maintain-aspect-ratio TRUE;  display "USEDEF";  valid_file "T";
%width 3.0104in;  height 1.8248in;  depth 0pt;  original-width 8.0332in;
%original-height 4.8568in;  cropleft "0";  croptop "1";  cropright "1";
%cropbottom "0";  tempfilename 'P1ADMA0H.bmp';tempfile-properties "XPR";}}}%
%BeginExpansion
%EndExpansion

\section{Analysis of Data from OPA}

We analyzed data from 12 experiments conducted over the last thirty
years \cite{vah}\cite{tak}-\cite{hir}. \ We
obtained values of $F_{T}$ and squeezing/antisqueezing from plots or from the
text, and estimated errors as much as possible. The most recent data is shown
in Figure 2 \cite{vah}. \ They fit their squeezing data $(S_{-}$ and $S_{+})$
to the model, actually including a small correction for the phase uncertainty,
with excellent agreement. \ To get $F_{T}$ \ from their data, we used the
equation from the OPA\ model in terms of the arctangent in Eq. \ref{ftgen}. \ On
the other hand, \ for the data in Figure 3 we could do calculations of
$F_{T}$ graphically from $S_{-}$ and $S_{+}$ and from $\Delta\theta,$ from
their plots. \ Thus, we can compare the two methods. \ (Some papers plot the
squeezing vs. time shown in Figure 2 as squeezing versus phase difference. \ We
treat both types of plots in the same manner with an assumed equivalence
between time and phase change that corresponds to the rate at which a mirror
is moved in degrees/second.) \ In about half the papers, we could compare the
two methods and found they agreed to within about $  \pm 8\%$ rms. \ When we could use
both methods to compute $F_{T}$, we used the average in our plots. \ For
Vahlbruch we took points at three power levels (x= 0.8, 0.3, 0.1) \cite{vah}.
For all other publications, we had only one power level.

\section{Interpretation of Squeezing and Observation Time}

To compare the results of the QI and the OPA data requires the assumption that
the squeezing in the OPA analysis is equivalent to the squeezing in the QI
analysis as discussed by Marecki \cite{marecki1}. \ In the OPA\ case, the
squeezing depends on the phase difference $\theta$ between the local
oscillator LO and the squeezed light SQZ, while in the QI analysis the squeezing
depends on the phase change $\omega_{o}t_{o}$ occurring during the observation
time. In the equation for $R$, which expresses the QI,
$t_{o}$ is the width of the Gaussian time sampling function and $\omega_{o}$
is the center frequency in radians/sec of the frequency sampling function. From Marecki's derivation, one would assume that this corresponds to the center frequency of the laser probe. However, once we are making measurements using BHD, where the detection is based on the interference between the LO and the squeezed light, the spectrum analyzer's output at a frequency $\omega$ is a quadrature noise measurement of the optical field at frequency $\omega_{LO} + \omega$.  Consequently the appropriate expression for the BHD phase change in radians during a measurement lasting a time $t_{o}$ is given by the product $\omega t_{o}$ where $\omega$ is the BHD measurement frequency. If we observe for a time interval
$M$, which equals the period of the squeezing $S(\theta,x,\omega)$, then the
phase change is $M\omega=\pi$. \ Therefore\ $\omega t=\pi(t/M)$,
where $t$ is the observation time. Defining the fractional observation time
$F_{T}=t/M$, we conclude that $\omega t=$ $\pi F_{T}$. \ Thus for a Gaussian
time sampling function we have%
\begin{equation}
R(F_{T})=10Log_{10}[Erf(\sqrt{2}\pi F_{T})] \label{ftg}%
\end{equation}

On the other hand, Marecki \cite{marecki1} just identified $\omega_{o}t_{o}%
$\ = $\tau$\ as the fraction of the period $F_{T}$ in which squeezing
occurred, omitted the factor of $\pi$, and also had an additional factor
of 2, thus obtaining $R(F_{T})= \\ 10Log_{10}[Erf(2\sqrt{2}F_{T})].$

For the squared Lorentzian time function we obtained%
\begin{equation}
R(F_{T})=10Log_{10}(1-e^{-2\pi F_{T}}) \label{loreq}%
\end{equation}
whereas Marecki did not have the factor $\pi$ in the exponent.

Note that in the derivation of the QI, $\omega t_{o}$ is
the phase change during an observation of the variance of a quadrature
component. \ Nothing is said in the derivation about whether the field is
squeezed or not. \ The QI appears to place a bound on the
variance for this phase change for any quadrature component, squeezed or not,
during the observation time. \ Assuming one physically can observe the field
only when it is squeezed, then we should obtain the value for $R(F_{T})$ as
restricted by the QI. \

\section{Comparison of QI Predictions\ and OPA\ Data}

If we assume that we are observing the variance during half of the period
then $F_{T}=0.5,$ and the QI gives a value of $R(F_{T})$
\ whose absolute value could not be exceeded with the maximum possible
squeezing during the half period. \ Similarly, if we are observing for the
entire period, then $F_{T}=1.$ \ By our understanding, the longer we observe,
the more likely we will have regions of variance that are above the vacuum
level and the bigger $R$ will be. \ For the shortest times, we can have the
most squeezing.

\begin{figure}[ptb]%
\centering
\includegraphics[
height=2.5962in,
width=3.300in
]%
{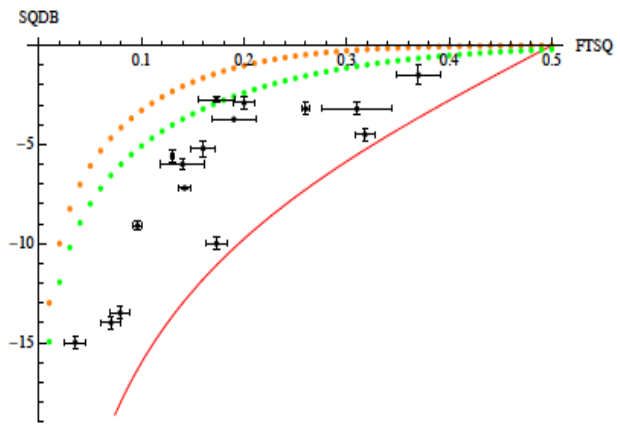}%
\caption{Squeezing $R$ in dB versus $F_{T},$ the fraction of observation period
that is squeezed, for a Gaussian time function. The top dotted curve is derived as $10Log_{10}[Erf(\sqrt{2}\pi
F_{T})$ ]. \ The middle dotted curve is calculated directly from equations in Reference 9. \ The
solid bottom curve is from the ideal OPA model. Experimental points are shown
with error bars. }%
\label{11}%
\end{figure}

In Figure \ref{11} for a Gaussian time function, we have plotted our result
for $R(F_{T})$ (Eq. \ref{ftg}, top dotted curve), and Marecki's result
$R(F_{T}) = 10Log_{10}[Erf(2\sqrt{2}F_{T}]$ (middle dotted curve), and $R(F_{T})$ for an ideal OPA
 (Eq. \ref{ftideal}, bottom solid curve). The QI is very restrictive; the degree of squeezing obtained in the experiments is greater than that allowed by either form of the QI for all but one experimental point. All data are consistent with the ideal OPA model.%

%TCIMACRO{\FRAME{ftbpFU}{3.8649in}{2.5962in}{0pt}{\Qcb{R dB of squeezing versus
%$F_{T},$ the fraction of observation period that is squeezed. The top dotted
%curve is derived as 10Log[erf($\sqrt{2}\pi F_{T})$ ]. \ The middle dotted
%curve is directly from Marecki's paper. \ The solid bottom curve is from the
%ideal OPA model.. Experimental points are shown with error bars. }}{\Qlb{11}%
%}{Figure}{\special{ language "Scientific Word";  type "GRAPHIC";
%maintain-aspect-ratio TRUE;  display "USEDEF";  valid_file "T";
%width 3.8649in;  height 2.5962in;  depth 0pt;  original-width 4.5939in;
%original-height 3.077in;  cropleft "0";  croptop "1";  cropright "1";
%cropbottom "0";  tempfilename 'P1ADMA0I.bmp';tempfile-properties "XPR";}}}%
%BeginExpansion
%EndExpansion
\
%TCIMACRO{\FRAME{ftbpFU}{3.5189in}{2.3973in}{0pt}{\Qcb{R db versus F$_{T}$ for
%a Lorentzian squared time function. \ Solid line, closest to the x-axis, is
%our result, which includes a factor of $\pi;$ middle dashed curve is Marecki's
%result, with no $\pi;$ and the thick dashed line is for an ideal OPA.}%
%}{\Qlb{loz}}{Figure}{\special{ language "Scientific Word";  type "GRAPHIC";
%display "USEDEF";  valid_file "T";  width 3.5189in;  height 2.3973in;
%depth 0pt;  original-width 5.0989in;  original-height 3.1324in;
%cropleft "0";  croptop "1";  cropright "1";  cropbottom "0";
%tempfilename 'P1ADMA0J.bmp';tempfile-properties "XPR";}}}%
%BeginExpansion
\begin{figure}[ptb]%
\centering
\includegraphics[
height=2.3973in,
width=3.300in
]%
{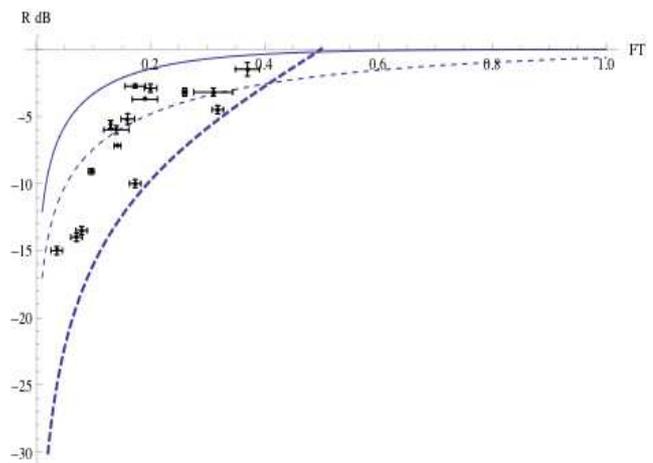}%
\caption{R dB versus $F_{T}$ for a Lorentzian squared time function. \ Solid
line, closest to the x-axis, is our result, which includes a factor of $\pi;$
middle dashed curve is Marecki's result with no $\pi;$ and the thick dashed
line is for an ideal OPA.}%
\label{loz}%
\end{figure}
%EndExpansion
The results for the squared Lorentzian time function were better than for the
Gaussian, as shown in Figure \ref{loz}. \ Almost all points violated Eq.
\ref{loreq}, but only about half the points violate Marecki's version of the
QI with no $\pi$ (middle dashed curve).

\begin{figure}[ptb]%
\centering
\includegraphics[
height=2.252in,
width=3.300in
]%
{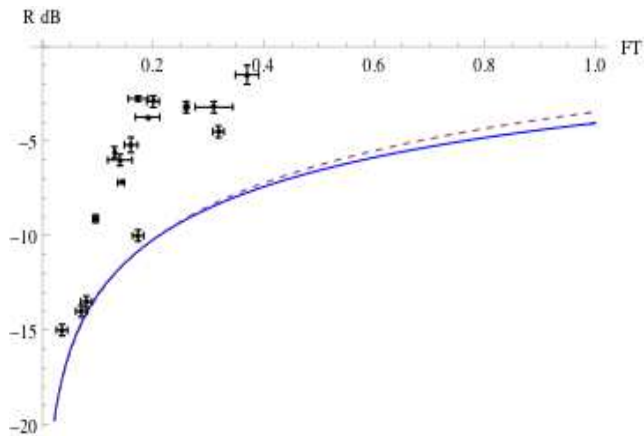}%
\caption{$R$ dB vesus $F_{T}$, showing the best fit for the Lorentzian (solid
curve) and the Gaussian (dashed curve) time functions. For the Lorentzian, the
arguments are (1/3$\pi)$ (or 1/3 for Marecki) of the theoretical values. For
the Gaussian, the arguments are (1/4$\pi)$ (or 1/4 for Marecki).}%
\label{yyy}%
\end{figure}

One phenomenological approach to understanding the disagreement between the data
and the QI is to try reducing the argument in the equations to improve the
agreement of the QI prediction with the data. As the arguments in the error function and the exponential decrease, the agreement does improve. Fitting the
functional forms to the data gives the plots as shown in Figure\ \ref{yyy}. \ No
points violate these best fits, but the significance of them is not clear.
Certainly for $F_{T}$ above about 0.3, they do not appear to be
sufficiently restrictive. They are not as restrictive as the ideal OPA curve.
%TCIMACRO{\FRAME{ftbpFU}{3.6201in}{2.252in}{0pt}{\Qcb{R db vesus F$_{T}$ ,
%showing the best fit for the Lorentzian (solid curve) and the Gaussian (dashed
%curve) time functions. For the Lorentzian, the arguments are (1/3$\pi)$ (or
%1/3 for Marecki) of the theoretical values. For the Gaussian the arguments are
%(1/4$\pi)$ (or 1/4$.$for Marecki).}}{\Qlb{yyy}}{Figure}%
%maintain-aspect-ratio TRUE;  display "USEDEF";  valid_file "T";
%width 3.6201in;  height 2.252in;  depth 0pt;  original-width 3.9557in;
%original-height 2.4509in;  cropleft "0";  croptop "1";  cropright "1";
%tempfilename 'jordan/P1ADMA0K.bmp';tempfile-properties "XPR";}}}%
%BeginExpansion
%EndExpansion

\begin{figure}[ptb]%
\centering
\includegraphics[
height=2.3419in,
width=3.300in
]%
{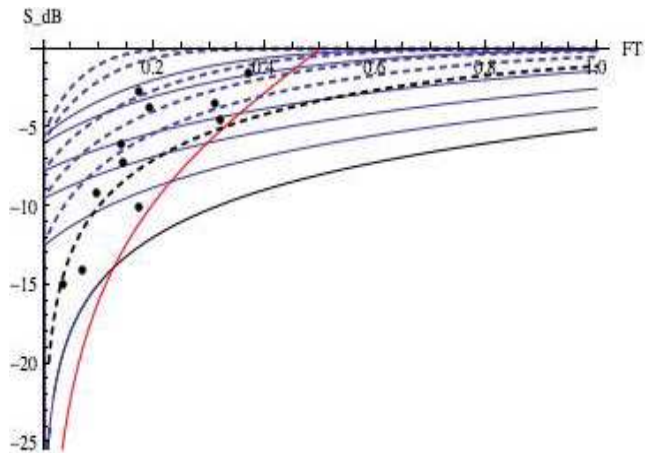}%
\caption{$R$ $dB$ $(=S_{-}$ $dB)$ versus $F_{T}$ for a symmetric
trapezoidal time function with a center region $T_{S}$ long and sides
$nT_{S}$ long.\ The dashed curves are for $n=0.001$ (most negative $R$), $ 0.2, 0.5,
1.0, 3.0, 5.0$ (nearest the origin). \ The solid curves are for Marecki's
result, which omitted the $\pi$. \ The solid curve crossing the other curves is
for the ideal OPA. \ }%
\label{xx}%
\end{figure}

We evaluated the variance for a symmetric trapezoidal window with a center
region $T_{S}$ long, and sloping sides that are each $nT_{S}$ long, normalized to 1.
The results (dashed curves) are displayed in Figure \ref{xx} \ for a range of
values of $n$ from $0.001$ (most negative black curve, and $f(t)$ is almost a
square window) to $5.0$ (dashed curve nearest the origin) which corresponds to
a nearly triangular window. \ The solid curves are for the same $n $ values, but
a factor of $\pi$ has been omitted in the argument in agreement with Marecki. The ideal
OPA bound is the solid curve crossing all other curves. As the window
becomes more triangular, the curves are less restrictive on squeezing and do
not agree with the data. \ Only the curve for the nearly square window
($n=0.001$) without the $\pi$ is not inconsistent with all the data. Yet this
curve clearly fails to be sufficiently restrictive for values of $F_{T}$
greater than about 0.3 and predicts squeezing exceeding that allowed by an ideal OPA. \ Mathematically this nearly rectangular window
is on the edge of instability, especially for low values of the power, and as
$n$ decreases further, this window becomes a square window for which the limit
is $R=10Log_{10}0$. \ %

%TCIMACRO{\FRAME{ftbpFU}{3.7974in}{2.3419in}{0pt}{\Qcb{Squeezing $R$ $dB$
%$(=S_{-}dB)$ versus $F_{T}$ for a symmetric trapezoidal time function, with a
%center region $T_{S}$ long, with sides $nT_{S}$ long.\ The dashed curves are
%for n=0.001 (most negative R), 0.2, 0.5, 1.0, 3.0, 5.0 (nearest the origin).
%\ The solid curves are for the Marecki's result, which omits the $\pi$. \ The
%solid curve crossing the other curves is for the ideal OPA. \ }}{\Qlb{xx}%
%}{squeezingmodel.jpg}{\special{ language "Scientific Word";  type "GRAPHIC";
%width 3.7974in;  height 2.3419in;  depth 0pt;  original-width 3.7498in;
%original-height 2.3021in;  cropleft "0";  croptop "1";  cropright "1";
%cropbottom "0";  filename 'squeezingModel.jpg';file-properties "XNPEU";}}}%
%BeginExpansion
%EndExpansion
Clearly the form of the time window is very important, yet for all forms
examined, the resultant QI do not appear to have the right
functional form or the numerical values one might expect for a QI applicable to the experimental data.

\section{Discussion and Conclusion}

The mathematics of the derivation of the Quantum Inequality appear sound, and
the model for the OPA data has been experimentally validated, and it is
consistent with all the data we examined. \ Yet the QI and OPA\ model do not
appear to be consistent with each other. The QI was violated by all the data
points for the Gaussian time window and about half the points for the
Lorentzian squared time window. \ Although other windows may show improved
results, these inconsistencies suggest a deeper problem.  

The model for the ideal OPA gave the best results: no data exceeded its maximum squeezing, yet the most recently published data came close.  It also predicted that the maximum duration of squeezed light does not exceed 1/2 the cycle, which agreed with the data, yet was not predicted by the QI.  Nevertheless, the ideal OPA is a model that is an approximation with limitations.

One of the issues mentioned was the potential conflict between specifying a
time function $f(t)$ and an independent frequency function $\mu(\omega)$. \ To
explore this effect, we did a calculation assuming the Gaussian time function
and a frequency function $\mu(\omega)$ which was given by the Fourier
transform of the time function. \ \ Explicit calculation showed that the
resulting expression for the QI was similar to that obtained
without explicitly giving the precise form of the frequency window. \ Although this conflict is real, it does not appear to be responsible for the
systemic disagreement seen between the QI and the OPA\ data.

Another possible issue might be the frequency dependence of the measured
squeezing for the OPA data as predicted by Eq. 13. The output beam
of a OPA has the Lorentzian squeezing spectrum with center frequency $\omega$ and
halfwidth $\gamma$\ that depends on the properties of the resonant cavity.
Experimental data are typically taken with a phase $\theta$ and frequency which give the
maximum squeezing. \ Since the QI correlates the fraction of
time the signal is squeezed with the dB of squeezing, we may need to account
for the change in $F_{T}$ for frequencies away from the sideband used for the
measurement. We can compute an "effective" duration of squeezing $F_{TE}(x)$
which is weighted by integrating the frequency over the variance of the
squeezed vacuum. The behavior of $F_{TE}(x)$ will depend on the
range over which we integrate and the halfwidth. This integration will
increase the effective size of the $F_{TE}(x),$ essentially moving all data
points to the right, making the disagreement between data and the QI worse, so
this is not the explanation.

Another critical issue concerns the nature of the assumed measurement in the
derivation of the Quantum Inequality. The assumption is that a measurement
of the energy density will be made that lasts a fraction of a cycle of oscillation of
the electromagnetic field of the laser being employed.  On the other hand, to make an accurate
measurement using BHD requires observation for a number of cycles.  It does not
appear possible to make a good measurement of the squeezing if
observation is for a fraction of a cycle.  The measurement of the energy density in
the OPA method is actually done over many cycles. For a fixed phase
difference between the LO and the vacuum signal SQZ, the balanced homodyne
detection automatically selects the corresponding energy output which is
measured continuously over as many cycles of the laser light as desired,
ensuring significant accuracy. On the other hand, no corresponding mechanism appears to be
available for the measurement assumed to occur in the derivation of the
QI. Thus, there may be an inconsistency
between the measurement assumed in the derivation of the QI and the
measurement method of the OPA.  Marecki addressed this issue in an analysis of the BHD method, stressing that in the theory of the QI all operators are restricted to the frequency $\omega_{LO}$ of the LO and time was $2\pi /\omega_{LO}$ periodic and therefore $\omega_{LO}t<2\pi$ for all times \cite{marecki2}. It is not clear if there is an inconsistency and if it is responsible for the disagreement between the QI and the experimental data.

The choice of window function is probably the most significant factor when applying the QI to
real data. \ Mathematically, the choice of a window function is simple.
\ However, when comparing theory to data, it is not clear what window function
is actually appropriate for the experiment being done  even though the choice dominates the restrictions due to the QI.
In addition, Heisenberg and Bohr maintained that measured fields were averages over space-time volumes, whereas Marecki (Eq. A.6) and Ford (Eq. 1) only have a time average.  

\ This work represents the only comparison to date of experimental data to the theory of a
QI. \ Hopefully, the conundrum of the disagreement between the QI
and the OPA measurements will be resolved more fully in the future with interdisciplinary collaborations and more
experiments, and more detailed theoretical derivations. \ Our results highlight
the subtleties that can be implicit in theoretical derivations of QI, particularly in the proposed measurement process. Ideally, an unambiguous experimental procedure could be associated with the theoretical derivations. These issues
may also affect the applicability of the QI that have been
proposed for other situations.

\appendix
\section{Derivation of Marecki's Quantum Inequality}
\setcounter{equation}{0}
\renewcommand{\theequation}{A.\arabic{equation}}

Marecki [9] derives a Quantum Inequality for squeezed light and squeezed vacuum
following the general approach of Fewster and Teo \cite{fenster} and Pfenning
\cite{pfen}. We briefly describe his derivation to clarify the comparisons to
the OPA data. Marecki defines the operator variance of the normally ordered
electric field$\ \Delta E^{2}(x,t)$:%
\begin{equation}
\Delta E^{2}(x,t)=E^{2}(x,t)-<E^{2}(x,t)>_{vac}%
\end{equation}
and considers a time sampling of the field squared%
\begin{equation}
\Delta=\int_{-\infty}^{+\infty}dtf(t)\Delta E^{2}(x,t)
\end{equation}

where
\begin{equation}
1=\int_{-\infty}^{+\infty}dtf(t)
\end{equation}

He also mentions the possibility of including a frequency sampling function
$\mu_{p}=$ $\mu(\omega_{p}-\omega_{0})$, peaked at $\omega_{0}$, that reflects the frequency response of
the apparatus measuring the variance. \ Since it is necessary to use a
frequency sampling function to get finite results for $\Delta,$ we will
include it in our derivations. \ However, we note that there is a potential
consistency issue using an independently selected frequency sampling function
since the time sampling function $f(t)$ implies a frequency selection
determined by its Fourier transform. \ Using the Coulomb gauge, the vector
potential is
\begin{multline}
A_{i}(\mathbf{x},t)=\frac{1}{\sqrt{(2\pi)^{3}}}\int\frac{d^{3}k}{\sqrt
{2\omega_{k}}}\times \\
\sum_{\alpha=1,2}\mathbf{e}_{i}^{\alpha}(\mathbf{k})\{a_{\alpha
}^{\dag}(\mathbf{k})e^{ikx}+a_{\alpha}(\mathbf{k})e^{-ikx}\}
\end{multline}
where $\omega_{k}=|k|$ and $\alpha$ denotes the two polarization states which
are normalized and orthogonal to $\mathbf{k}$. In the exponentials,
$kx=-\mathbf{k}\cdot\mathbf{x}+\omega t$ represents the scalar product. The
electric field operator is%
\begin{equation}
E_{i}=-\frac{\partial A_{i}}{\partial t}%
\end{equation}
The expectation value of the time sampled free vacuum field squared is
\begin{equation}
<E^{2}>_{vac}=\int_{-\infty}^{+\infty}dtf(t)<E^{2}(\mathbf{x},t)>_{vac}
\end{equation}

\begin{multline}
=\frac{1}{2(2\pi)^{2}}\int\mu_{k}d^{3}k\mu_{p}d^{3}p\sqrt{\omega_{k}%
\omega_{p}}2\pi f_{FT}(\omega_{p}-\omega_{k})\times \\
\sum_{\alpha,\beta
=1,2}\mathbf{e}_{i}^{\alpha}(\mathbf{k})\mathbf{e}_{i}^{\beta}(\mathbf{p}%
)\delta_{\alpha\beta}\delta(\mathbf{p}-\mathbf{k})
\end{multline}

where we have used the commutator $[a_{\alpha}(\mathbf{p}),a_{\beta}^{\dag
}(\mathbf{k})]=\delta_{\alpha\beta}\delta(\mathbf{p}-\mathbf{k})$ and
included the frequency function.  The Fourier transform of the time sampling function $f(t)$ is defined as%
\begin{equation}
f_{FT}(\omega)=\frac{1}{2\pi}\int_{-\infty}^{+\infty}dtf(t)e^{-i\omega t}%
\end{equation}
Integrating Eq. A.7 over k, using the unity normalization of the polarization
vectors $%
%TCIMACRO{\dsum \limits_{i}}%
%BeginExpansion
{\displaystyle\sum\limits_{i}}
%EndExpansion
\mathbf{e}_{i}^{\alpha}(\mathbf{k})\mathbf{e}_{i}^{\alpha}(\mathbf{p})=1$ ,
and that $f_{FT}(0)=1/2\pi$ because of the $f(t)$ nomalization, gives
\begin{equation}
<E^{2}>_{vac}=\frac{1}{(2\pi)^{3}}\int(\mu_{p}^{2}\omega_{p})d^{3}p
\end{equation}

Substituting this result into the expression for the variance $\Delta$ gives,
after integration over time,
\begin{multline}
\Delta=\frac{1}{2(2\pi)^{2}}\int d^{3}kd^{3}p\mu_{k}\mu_{p}\sqrt{\omega
_{k}\omega_{p}}\times\\ \sum_{\alpha,\beta=1,2}\mathbf{e}_{i}^{\alpha}%
(\mathbf{k})\mathbf{e}_{i}^{\beta}(\mathbf{p})\{a_{\alpha}^{\dag}%
(\mathbf{k})a_{\beta}(\mathbf{p})e^{i(-\mathbf{k}+\mathbf{p})\mathbf{x}}%
f_{FT}(\omega_{p}-\omega_{k})\\
-a_{\alpha}(\mathbf{k})a_{\beta}(\mathbf{p})e^{i(\mathbf{k}+\mathbf{p}%
)\mathbf{x}}f_{FT}(\omega_{p}+\omega_{k})+HC\}
\end{multline}
\linebreak where HC is the Hermitian conjugate. \ To derive a quantum
inequality, Marecki defines a vector operator $B_{i}(\omega)$ and computes
the integral over frequency of the norm of $\mathbf{B}$ which has to be positive%
\begin{equation}
\int_{0}^{\infty}d\omega B_{i}^{\dag}(\omega)B_{i}(\omega)>0
\end{equation}
We choose
\begin{multline}
B_{i}(\omega)=  \frac{1}{\sqrt{2\pi^{2}}}\int d^{3}p\sqrt{\omega_{p}} \times \\
\sum_{\alpha,\beta=1,2}\mathbf{e}_{i}^{\alpha}(\mathbf{p})\{a_{\alpha
}(p)(f^{1/2})_{FT}^{\ast}(\omega-\omega_{p})e^{i\mathbf{px}}- \\ a_{\alpha}^{\dag
}(p)(f^{1/2})_{FT}^{\ast}(\omega+\omega_{p})e^{-i\mathbf{px}}\}
\end{multline}
and substitute this into Eq. A.11, and use the result in Eq. A.10. After taking the expectation value with respect to state A, we obtain Eq. 4 in Section 2. Note that in Eq. A.12, 
$(f^{1/2})_{FT}^{\ast}(\omega - \omega_{p})$ means the complex conjugate of the Fourier transform of the square root of $f(t)$.  

\begin{acknowledgement} We are very grateful to Peter Milonni and Larry Ford for helpful discussions and comments. We would like to thank the Institute for Advanced Studies At Austin and H. E. Puthoff for supporting this work.
\end{acknowledgement}

\end{document}